\documentclass
[superscriptaddress,secnumarabic,amssymb,amsmath,nobibnotes,aps,prd,showkeys,nofootinbib,nopacsnumber,onecolumn,10pt]{revtex4-2}%
\usepackage{graphicx}
\usepackage{bm}
\usepackage{amsmath}
\usepackage{amsfonts}
\usepackage{amssymb}%
\usepackage[table]{xcolor}
\usepackage{color}
\usepackage{booktabs}
\usepackage{float}
\providecommand{\U}[1]{\protect\rule{.1in}{.1in}}

\newcommand{\be}{\begin{equation}}
\newcommand{\ee}{\end{equation}}

\newcommand{\mincir}{\raise
-3.truept\hbox{\rlap{\hbox{$\sim$}}\raise4.truept\hbox{$<$}\ }}
\newcommand{\magcir}{\raise
-3.truept\hbox{\rlap{\hbox{$\sim$}}\raise4.truept\hbox{$>$}\ }}

\begin{document}

\title{Constraints on Chiral-Quintom dark energy after DESI DR2 and impact on unifying dark energy with inflation}

\author{Andronikos Paliathanasis}
\email{anpaliat@phys.uoa.gr}
\affiliation{Institute of Systems Science, Durban University of
Technology, Durban 4000, South Africa.}
\affiliation{Departamento de Matem\`{a}ticas, Universidad Cat\`{o}lica del Norte, Avda.
Angamos 0610, Casilla 1280 Antofagasta, Chile.}
\affiliation{National Institute for Theoretical and Computational Sciences (NITheCS), South Africa.}

\author{Tommaso Mengoni}
\email{tommaso.mengoni@unicam.it}
\affiliation{University of Camerino, Via Madonna delle
Carceri, Camerino, 62032, Italy.}
\affiliation{INAF - Osservatorio Astronomico di Brera, Milano, Italy.}
\affiliation{Istituto Nazionale di Fisica Nucleare (INFN), Sezione di Perugia, Perugia,
06123, Italy.}

\author{Genly Leon}
\email{genly.leon@ucn.cl}
\affiliation{Institute of Systems Science, Durban University of
Technology, Durban 4000, South Africa.}
\affiliation{Departamento de Matem\`{a}ticas, Universidad Cat\`{o}lica del Norte, Avda.
Angamos 0610, Casilla 1280 Antofagasta, Chile.}

\author{Orlando Luongo}
\email{orlando.luongo@unicam.it}
\affiliation{University of Camerino, Via Madonna delle
Carceri, Camerino, 62032, Italy.}
\affiliation{INAF - Osservatorio Astronomico di Brera, Milano, Italy.}
\affiliation{Istituto Nazionale di Fisica Nucleare (INFN), Sezione di Perugia, Perugia,
06123, Italy.}
\affiliation{Department of Nanoscale Science and Engineering, University at Albany-SUNY,
Albany, New York 12222, USA.}
\affiliation{Al-Farabi Kazakh National University, Al-Farabi av. 71, 050040 Almaty, Kazakhstan.}

\begin{abstract}
The recent data release DR2 from the Dark Energy Spectroscopic Instrument (DESI) has reinforced compelling evidence supporting the dynamical nature of dark energy. In this respect, we here explore a two-scalar field cosmological model, dubbed  Chiral-Quintom paradigm, that allows for a unified description of early- and late-time cosmic accelerations, namely inflation and dark energy, respectively. Moreover, we show that it provides a mechanism to cross the phantom divide without instabilities. To do so, we first focus on scenarios where the scalar fields evolve under an exponential potential, leading to distinct cosmological behaviors, including tracking solutions and slow-roll or hyperbolic inflation. Afterwards, by considering a nonlinear potential and a mixing in the kinetic sector, we show that the model can also describe the matter-dominated era, offering a potential unification of the dark sector. Accordingly, we place bounds on this double-field paradigm by employing recent low-redshift observational data, including supernovae from the Pantheon+ and Union3.0 compilations, observational Hubble parameter measurements, and baryon acoustic oscillation data from the latest DESI release. By means of the Akaike information criterion, the standard $\Lambda$CDM scenario and the $w_0w_a$CDM model are thus  compared with the Chiral-Quintom approach, showing that, in principle, multiple-field dynamics does not seem to be ruled out, ultimately providing a flexible framework for describing late-time dynamics.
\end{abstract}


\maketitle

\section{Introduction}

The second DESI data release, DR2~\cite{des5}, updates the previous results, further reinforcing the possible evidence for evolving dark energy and providing a mild indication in favor of the $w_0w_a$CDM model, thus challenging the $\Lambda$CDM paradigm and the corresponding cosmological constant~\cite{des6, des4}. Since then, several works have analyzed these data points to infer the dark energy nature, see  e.g.~\cite{d01,d02,d03,d04,d05,d06,d07,d08,d09,d10,d11,d12,d13,d14,d15,d16,d17,d18,d19,d20,d21,d22,d23,d25}, even though criticisms against the DESI results have also been raised, see e.g.~\cite{Luongo:2024fww,d25,Yang:2021oxc,Colgain:2025fct,Colgain:2025nzf}.

Immediately, scalar field models have been employed to explain the DESI results~\cite{sf1,sf2,sf3,sf4,sf5,sf6,sf7,sf8,sf9}, as they turn out to be the simplest addition to Einstein's gravity beyond the cosmological constant, since their dynamics may describe both early and late times, see Refs.~\cite{guth,Ratra,in1}.

A great advantage of scalar field theories is that they include modifications of the Einstein–Hilbert action~\cite{sot1}, and general forms such as Horndeski gravity~\cite{hornd} and other scalar–tensor models~\cite{Fujii:2003pa}, which can be mapped into higher-order actions~\cite{Nojiri:2010wj}. In this respect, multi–scalar field theories are endowed with a mechanism that enables dark energy to cross the phantom divide line without introducing ghost instabilities~\cite{qq1,qq2,qq3,qq4,qq5}. 

In addition to what has been described above, it appears natural to wonder whether the role of multiple scalar fields can or cannot be a mere speculation or if interactions among multiple fields may lead to observational signatures, for example  acting as sources for quantum effects \cite{Lyth:2001nq}, reheating background \cite{Kofman:1994rk}, interacting sector between dark energy and dark matter \cite{Amendola:1999er} and so on. 

Motivated by the above, we consider a two-scalar field cosmological model in order to unify early- and late- time  accelerations. To do so, we employ a direct extension of the \emph{quintom model} \cite{qq1,qq2,qq3,qq4,qq5} that is the among most studied multi–scalar-field framework for dark energy, here differing as we introduce a non-constant function in the kinetic sector, giving rise to a non-minimal picture, whose coupling constant is, moreover, chosen in agreement with possible unified inflationary–dark energy models~\cite{ns03}. In this respect, we baptize our scenario with the indicative name of \emph{Chiral-Quintom} scalar-field framework, invoking a unification scheme along the entire cosmic history. In this respect, indeed, this multi-scalar framework has been shown to yield a new phase of cosmic acceleration distinct from the conventional slow-roll regime, known as \emph{hyperbolic inflation} \cite{hyp1}. The novelty of hyperbolic inflation lies in the fact that the initial and final conditions of inflation can differ, with curvature perturbations depending on the number of e-folds \cite{ch1}, and non-Gaussianities appearing in the power spectrum \cite{ch2,ch4}. When one of the scalar fields, $\psi$, becomes constant, the model recovers the well-known tracking solution of quintessence. In this limit, the theory can also reproduce the standard slow-roll inflationary scenario \cite{ns03,ancqg}.

Moving on from this approach, the exponential potential $V(\phi)=V_{0}e^{\lambda\phi}$, extensively studied for the description of hyperbolic inflation, is here considered. In particular, from the Chiral theory, the cosmological solutions depend on two parameters, $\lambda$ and $\kappa$, where $\kappa$ is a free parameter of the model. Indeed, as discussed previously, the tracking solution depends solely on the parameter $\lambda$, for which only the scalar field $\phi$ contributes to the cosmic fluid. This solution describes acceleration for $\lvert \lambda\rvert <\sqrt{2}$. On the other hand, both scalars contribute to the cosmic fluid for the second asymptotic solution, which describes inflation for $-\frac{2\kappa}{\kappa+\lambda}<-\frac{4}{3}$. The solution tends asymptotically to the de Sitter universe for $\kappa\gg\lambda$. For initial conditions around this second solution, the trajectories are periodic, permitting an oscillatory behavior of the dark energy equation of state parameter, see~\cite{ancqg}.

Nonetheless, a nonlinear potential that mixes the two scalar fields in the kinetic sector turns the model into one capable of describing the matter-dominated epoch and, hence, of unifying the dark sector of the universe \cite{anuni,orly1}. The importance of this approach lies on unifying inflation and dark energy, as stated above \cite{sa1}. Recently, the gravitational waves of the background spacetime of this multi–scalar-field model were studied in \cite{orly2}, while other generalizations have been explored in Refs.  \cite{cq1,cq2,cq3,cq4,cq5,cq6,cq7}.

In this work, we focus on constraining the Chiral--Quintom model at the background level, 
aiming to determine whether its free parameters can be  bounded by modern 
late-time cosmological observations, including the most recent DESI DR2 measurements. 
In so doing, after presenting the unified multi-scalar framework and discussing its stability properties, 
we adopt the exponential potential as an asymptotic limit of more general potentials and 
numerically reconstruct the Hubble function $H(z)$, since for generic values of $(\kappa,\lambda)$ 
no closed-form expression is available. To do so, we recast the cosmological field equations 
into the Hubble-normalized dynamical variables $(x,y,\xi,\lambda)$, imposing algebraic 
constraints along with physically motivated conditions on the parameter space. Accordingly, the numerical background evolution is then implemented through the \textsc{Cobaya} framework, employing a Metropolis--Hastings Monte Carlo Markov chain (MCMC) analysis and 
analyzing the resulting chains through the free-available \textsc{GetDist} package. We explore different 
combinations of late-time probes, namely type Ia Supernovae (SNe~Ia) from the Pantheon+ and Union~3.0 
catalogs, observational Hubble parameter measurements from Cosmic Chronometers (OHD), 
and baryon acoustic oscillation (BAO) distances from the DESI DR2 release. To assess the statistical performance of the model, we compare the Chiral--Quintom 
framework with both the standard $\Lambda$CDM scenario and the Chevallier-Polarski-Linder (CPL) parameterization\footnote{More precisely, the parametrization is currently known as $w_0w_a$CDM model, to enable the phase space of free parameters, namely $w_0--w_a$ to lie on phantom regimes too, that in the original formulation, namely the CPL model, was not fully assumed. The parametrization is probably far from being  a definitive paradigm, albeit it seems to well behave while compared with the most recent cosmic data. We here indistinguishably refer to it as either  CPL parametrization or $w_0w_a$CDM model.}. 
Following Akaike’s prescription, we compute the AIC for each dataset combination to 
quantify the relative preference among competing cosmological models. The combined SNe~Ia+OHD+DESI datasets tend to reward the additional flexibility 
introduced by the two-field dynamics, say Pantheon+ mildly favors the Chiral--Quintom scenario, 
while Union~3.0 provides strong evidence in its favor with $\Delta\mathrm{AIC}< -4$. 
The reconstruction of $w_{\mathrm{CQ}}(z)$ further confirms that the present expansion is 
driven by the asymptotic regime in which both scalar fields contribute to the cosmic fluid, 
yielding a dynamical dark energy component capable of crossing the phantom divide without evident instabilities. In addition, our analysis seems to show that multi-field dark energy 
models can perform better with respect to the $\Lambda$CDM paradigm, while describing the dynamics even at late-times data, in mild agreement with the 
trend suggested by the DESI mission. We thus end up our paper with the suggestion that such a kind of approaches, namely multi-field dark energy, may represent a naive technique to better describe the universe expansion history, noticing that dark energy may be due to the interactions among spectator fields. 

The structure of the paper is optimized as follows. In Sect. \ref{sec2} we present our theoretical model. The main analysis of this study is then presented in Sect. \ref{sec3}, where we examine the Chiral-Quintom model as dark energy candidate and, after constraining it with cosmic data points, we draw our conclusions and physical perspectives, while finally in Sect. \ref{sec4} we draw our final outlooks and perspectives.

\section{The Chiral-Quintom model}\label{sec2}

For the description of the gravitational field, inspired from the nonlinear
$\sigma~$model \cite{sigm0} we consider the Action Integral
\begin{equation}
S=S_{EH}+S_{CQ}+S_{m} \label{qq.01}%
\end{equation}
where $S_{EH}$ is the Einstein Hilbert Action, $S_{CQ}$ is the Action
Integral for the Chiral-Quintom model, which describes a two-scalar field
theory, and $S_{m}$ is the Action Integral for the matter source, which we
consider as an ideal gas with energy density $\rho_{m}$. Therefore,
the latter Action Integrals are defined as follows \cite{ns01,ns02,ns03,atr7}%
\begin{subequations}
    \begin{align}
S_{EH}  &  =\int d^{4}x\sqrt{-g}R,\label{qq.02}\\
S_{CQ}  &  =-\int d^{4}x\sqrt{-g}\left(  \frac{1}{2}g^{\mu\nu}\nabla_{\mu}%
\phi\nabla_{\nu}\phi+\frac{1}{2}g^{\mu\nu}e^{\kappa\phi}\nabla_{\mu}\psi
\nabla_{\nu}\psi+V\left(  \phi\right)  \right)  ,\label{qq.03}\\
S_{m}  &  =-\int d^{4}x\sqrt{-g}\rho_{m}. \label{qq.04}%
    \end{align}
\end{subequations}

For the Chiral-Quintom model we observe that the two scalar fields interact in
the kinetic term. In particular, the \textquotedblleft kinetic
energy\textquotedblright\ is defined by a two-dimensional metric of constant
curvature, that is, the hyperbolic plane. Parameter $\kappa$, plays an
important role in the dynamics of the model, because it provides the nonzero
interaction. In the limit $\kappa=0$, the theory is reduced to the usual
quintom model. For the model (\ref{qq.03}), in the limit $\kappa=0$, the
scalar field $\psi$, plays the role of a kination field, that is, it describes
a stiff fluid component.
This model, together with some generalized classes of two scalar field theories~\cite{sa1}, 
provides a consistent framework unifying inflation and dark energy within a single scheme, 
where the universe is accelerated by these nonminimally coupled fields.
Notably, the choice of the nonminimal coupling represents a natural way to connect the two cosmological epochs 
while preserving generality; indeed, we can still recover further scenarios, 
such as warm inflation or a unification scheme involving dark matter as well~\cite{orly2}.

Variation with respect to the metric of Eq. (\ref{qq.01}) leads to  Einstein's
field equations%
\begin{equation}
G_{\mu\nu}=T_{\mu\nu}^{CQ}+T_{\mu\nu}^{m}, \label{qq.05}%
\end{equation}
in which $G_{\mu\nu}\equiv R_{\mu\nu}-\frac{1}{2}Rg_{\mu\nu}$ is the Einstein
tensor, $T_{\mu\nu}^{m}$ is the energy momentum tensor for the matter field,
that is,%
\begin{equation}
T_{\mu\nu}^{m}=\left(  \rho_{m}+p_{m}\right)  u_{\mu}u_{\nu}+p_{m}g_{\mu\nu},
\label{qq.06}%
\end{equation}
and $T_{\mu\nu}^{CQ}$ is the energy momentum tensor for the Chiral-Quintom
model, that is
\begin{align}
T_{\mu\nu}^{CQ}    =\nabla_{\mu}\phi\nabla_{\nu}\phi-g_{\mu\nu}\left(
\frac{1}{2}g^{\kappa\lambda}\nabla_{\kappa}\phi\nabla_{\lambda}\phi+V\left(
\phi\right)  \right) +\frac{1}{2}\left(  e^{\kappa\phi}\nabla_{\mu}\psi\nabla_{\nu
}\psi-\frac{1}{2}g_{\mu\nu}\left(  g^{\kappa\lambda}\nabla_{\kappa}\psi
\nabla_{\lambda}\psi\right)  \right)  . \label{qq.07}%
\end{align}

Furthermore, the Bianchi Identity provides the continuous equations for the
scalar fields and the matter source, they are, $\nabla_{\nu}\left(
T_{~~~~~~~~}^{m~\mu\nu}\right)  =0$, and$~\nabla_{\nu}\left(  T^{CQ~\mu\nu
}\right)  =0$. The latter provides the \textquotedblleft
Klein-Gordon\textquotedblright\ equations$~$
\begin{subequations}
    \begin{align}
\frac{1}{\sqrt{-g}}\left(  \partial_{\mu}\left(  \sqrt{-g}g^{\mu\nu}%
\partial_{\nu}\phi\right)  \right)  -\frac{1}{2}\kappa e^{\kappa
\phi}\left(  g^{\mu\nu}\partial_{\mu}\psi\partial_{\nu}\psi\right)
+V_{,\phi}  &  =0,\label{qq.08}\\
\frac{1}{\sqrt{-g}}\left(  \partial_{\mu}\left(  \sqrt{-g}g^{\mu
\nu}e^{\kappa\phi}\partial_{\nu}\psi\right)  \right)   &  =0. \label{qq.09}%
\end{align}
\end{subequations}

Equation (\ref{qq.09}) states that the current~
$\Psi^{\mu}\left(  x^{\kappa}\right)  =\left(  \sqrt{-g}g^{\mu\nu}e^{\kappa
\phi}\partial_{\nu}\psi\right)$ 
is a conserved quantity, i.e., $\partial_{\mu}\Psi^{\mu}=0$~\cite{ns02}.

\subsection{The chiral-quintom model in homogeneous and isotropic cosmology}

For the background geometry, we consider a spatially flat FLRW metric with line
element
\begin{equation}
ds^{2}=-dt^{2}+a\left(  t\right)  ^{2}\left(  dx_{1}^{2}+dx_{2}^{2}+dx_{2}%
^{2}\right)  , \label{qq.10}%
\end{equation}
where $a\left(  t\right)  $ is the scale factor and for the comoving observer
$u^{\mu}=\delta_{t}^{\mu}$, the expansion rate is given as $\theta=\frac{1}%
{3}H$, where $H=\frac{\dot{a}}{a}$ is the Hubble function and $\dot{a}%
=\frac{da}{dt}$. Furthermore, we assume the energy-momentum tensor $T_{\mu\nu
}^{m}$ to describe homogeneous cold matter, i.e. $p_{m}=0,~\rho_{m}=\rho
_{m}\left(  t\right)  $, while for the scalar fields $\phi,\psi$ we consider
that they inherit the symmetries of the background spacetime (\ref{qq.10}),
that is, $\phi=\phi\left(  t\right)  $ and $\psi=\psi\left(  t\right)  $.

Hence, the gravitational field equations (\ref{qq.05}) for the spatially flat
FLRW geometry are \cite{ns02}%
\begin{subequations}
    \begin{align}
3H^{2}  &  =\left(  \frac{1}{2}\dot{\phi}^{2}+V\left(  \phi\right)  \right)
+\frac{1}{2}e^{\kappa\phi}\dot{\psi}^{2}+\rho_{m},\label{qq.11}\\
-\left(  2\dot{H}+3H^{2}\right)   &  =\left(  \frac{1}{2}\dot{\phi}%
^{2}-V\left(  \phi\right)  \right)  +\frac{1}{2}e^{\kappa\phi}\dot{\psi}^{2}.
\label{qq.12}%
    \end{align}
\end{subequations}

Moreover, the \textquotedblleft Klein-Gordon\textquotedblright\ relations, Eqs.
(\ref{qq.08}) and (\ref{qq.09}), become%
\begin{align}
\left(  \ddot{\phi}+3H\dot{\phi}\right)  +\frac{1}{2}\kappa e^{\kappa\phi}%
\dot{\psi}^{2}+V_{,\phi}  &  =0,\label{qq.13}\\
\ddot{\psi}+3H\dot{\psi}+\kappa\dot{\phi}\dot{\psi}  &  =0. \label{qq.14}%
\end{align}
The conserved quantity $\Psi_{\mu}\left(  x^{\kappa}\right)  $, becomes an
integration constant, that is, $\Psi_{0}=a^{3}e^{\kappa\phi}\dot{\psi}$.
Finally, the conservation equation for the cold matter provides $\rho_{m}%
=\rho_{m0}a^{-3}$, with the integration constant $\rho_{m0}$ to describe the
energy density of the cold matter at the present time.

With the application of the conservation of $\Psi_{0}$, the cosmological
field equations (\ref{qq.11}), (\ref{qq.12}) and (\ref{qq.13}) are simplified
as follows%
\begin{subequations}
    \begin{align}
3H^{2}-\frac{1}{2}\dot{\phi}^{2}-V\left(  \phi\right)  -\frac{1}{2}\Psi
_{0}^{2}a^{-6}e^{-\kappa\phi}-\rho_{m0}a^{-3}  &  =0\label{qq.15}\\
2\dot{H}+3H^{2}+\frac{1}{2}\dot{\phi}^{2}-V\left(  \phi\right)  +\frac{1}%
{2}\Psi_{0}^{2}a^{-6}e^{-\kappa\phi}  &  =0,\label{qq.16}\\
\left(  \ddot{\phi}+3H\dot{\phi}\right)  +\frac{1}{2}\kappa\Psi_{0}^{2}%
a^{-6}e^{-\kappa\phi}+V_{,\phi}  &  =0. \label{qq.17}%
\end{align}
\end{subequations}

The integrability properties of this model are investigated in detail in
\cite{chr1}. It was found that for $\lambda=\pm\frac{\sqrt{6}}{2}$ and
$\kappa=\mp\frac{3\sqrt{6}}{2}$, the cosmological field equations
(\ref{qq.15}), (\ref{qq.16}) and (\ref{qq.17}) form a Liouville integrable
dynamical system. Nevertheless, in the vacuum, integrability is achieved for
$\kappa=-\left(  \lambda+\sqrt{6}\right)  $. Furthermore, for other functional
forms of the potential, analytic solutions have been derived before in \cite{ns02}.

In the following, we investigate Chiral-Quintom model, with the exponential
potential, as a dark energy candidate. That is, we explore whether it can describe
the late-time cosmic acceleration phase of the universe. To perform such
analysis, we constrain the free parameters of the model with cosmological observations.

\section{Stability and observational Analysis}

\label{sec3}

Although for specific values of parameters $\lambda$ and $\kappa$ the field
equations are integrable, the Hubble function $H=H\left(  a\right)  $, can not be
expressed in closed-from. Thus, we proceed with the numerical reconstruction of
the Hubble function. To perform such analysis, we introduce the new
dimensionless variables within the Hubble-normalization approach%
\begin{equation}
x=\frac{\dot{\phi}}{\sqrt{6}H}~,\,\,~y=\frac{\sqrt{V\left(  \phi\right)  }}%
{\sqrt{3}H},\,\,\text{~}\xi=\frac{e^{\frac{\kappa}{2}}\dot{\psi}}{\sqrt{6}%
H},\,\,~\lambda=\frac{V_{,\phi}}{V},\,\,~\Omega_{m}=\frac{\rho_{m0}a^{-3}}{3H^{2}}.
\label{qq.18}%
\end{equation}

In terms of the new variables the field equations read \cite{ancqg}%
\begin{subequations}
    \begin{align}
-\left(  1+z\right)  \frac{dx}{dz}  &  =\frac{1}{2}\left(  3x^{3}-3x\left(
1+y^{2}-\xi^{2}\right)  +\sqrt{6}\left(  \kappa\xi^{2}-\lambda y^{2}\right)
\right)  ,\label{qq.19}\\
-\left(  1+z\right)  \frac{dy}{dz}  &  =\frac{1}{2}y\left(  3\left(
1-y^{2}+x^{2}+\xi^{2}\right)  +\sqrt{6}\lambda x\right)  ,\label{qq.20}\\
-\left(  1+z\right)  \frac{d\xi}{dz}  &  =\frac{1}{2}\xi\left(  3\left(
x^{2}+\xi^{2}-y^{2}-1\right)  -\sqrt{6}\kappa x\right)  ,\label{qq.21}\\
-\left(  1+z\right)  \frac{d\lambda}{dz}  &  =\sqrt{6}\lambda^{2}x\left(
\Gamma\left(  \lambda\right)  -1\right)  , \label{qq.22}%
    \end{align}
\end{subequations}

where $z$ is the redshift, i.e. $1+z=a^{-1}$, and with algebraic constraint%
\begin{equation}
\Omega_{m}=1-x^{2}-y^{2}-\xi^{2}. \label{qq.23}%
\end{equation}

Furthermore, the equation of state parameter for the scalar fields, which we
will consider as dark energy equation of state parameter, is expressed as
follows \cite{ancqg}%
\begin{equation}
w_{CQ}=\frac{x^{2}-y^{2}+\xi^{2}}{x^{2}+y^{2}+\xi^{2}}. \label{qq.23a}%
\end{equation}

We consider the exponential potential, from where it follows that $\lambda$ is
always a constant parameter. Moreover, we observe that the dynamical system is
invariant under the discrete transformations $y\rightarrow-y$, $\xi
\rightarrow-\xi$ and $\left(  x,\kappa,\lambda\right)  \rightarrow-\left(
x,\kappa,\lambda\right)  $.

In order to reconstruct the Hubble function, we solve numerically the field
equations (\ref{qq.19})-(\ref{qq.21}), with initial conditions at the present,
i.e. $z=0,$~$IC=\left(  x_{0},y_{0},\xi_{0}\right)  ^{T}$ and
\begin{equation}
y_{0}=\sqrt{1-\Omega_{m0}-x^{2}-\xi^{2}}. \label{qq.24}%
\end{equation}
The initial conditions are selected such that%
\begin{equation}
\operatorname{Re}\left(  y_{0}\right)  \geq0,~\operatorname{Im}\left(
y_{0}\right)  =0, \label{qq.25}%
\end{equation}
and $0\leq\Omega_{m0}\leq1$. The latter conditions mean that the dynamical
variables $x$ and $\xi$ lie on a unitary disk of radius $\left(  1-\Omega
_{m0}\right)  $.

Last but not least, variables $\kappa,\lambda~$\ are the additional free
parameters of the model that should be constrained from the observational data.

\subsection{Cosmological data}

The late-time cosmological observations that we consider in this work are the
Supernova (SNe~Ia), the Observational Hubble data (OHD) from the Cosmic
Chronometers and the Baryonic Acoustic Oscillations.

\begin{itemize}
\item[-] Pantheon+ (PP): This SNe~Ia catalog consists of 1701 light curves
of 1550 spectroscopically confirmed supernova events within the range
$10^{-3}<z<2.27~$\cite{pan}, which relate the observable distance modulus
$\mu^{obs}$ at observed redshifts. For a spatially flat FLRW geometry, the
theoretical distance modulus is derived from the Hubble functions as 
\begin{equation}
 \mu
^{th}=5\log D_{L}+25,   
\end{equation}
where $D_{L}$ is the luminosity distance~
\begin{equation}
D_{L}%
=\left(  1+z\right)  \int\frac{dz}{H\left(  z\right)  }.    
\end{equation}
We apply the PP data without the SH0ES Cepheid calibration.

\item[-] Union3.0 (U3): This is the most recent SNe~Ia catalog, which includes
2087 events within the same redshift range as the PP catalog \cite{union}.
There are 1363 common events between the two catalogs. The main difference
between the two catalogs is the approach applied for the analysis of the
direct photometric observations giving different observable values.

\item[-] Cosmic Chronometers (CC): Cosmic Chronometers are old galaxies,
passively evolving with synchronous stellar populations and similar cosmic
evolution \cite{co01}. The difference in the age of the galaxies at different
redshifts leads to direct values for the Hubble parameter. In this work, we
apply the 31 model independent direct measurements of the Hubble parameter
within the redshift range $0.09\leq z\leq1.965$~\cite{cc1}

\item[-] The DESI DR2: The dataset relates
observation values of the transverse comoving angular distance ratio, 
\begin{equation}
\frac{D_{M}}{r_{drag}}=\frac{\left(1+z\right)  ^{-1}}{r_{drag}}D_{L},    
\end{equation}
the
volume averaged distance ratio 
\begin{equation}
    \frac{D_V}{r_{\mathrm{drag}}}
= 
\frac{\bigl(D_L \, z / H(z) \bigr)^{1/3}}{r_{\mathrm{drag}}}.
\end{equation}
and the Hubble distance ratio
\begin{equation}
\frac{D_{H}}{r_{drag}}=\frac{1}{r_{drag}H}    
\end{equation}
at each of the seven redshift points, while $r_{drag}$ denotes the sound horizon at the drag epoch.
\end{itemize}

\subsection{Methodology and Priors}

For the observational constraints and the parameter estimation, we apply the
Bayesian inference COBAYA\footnote{https://cobaya.readthedocs.io/}
\cite{cob1,cob2} with the MCMC sampler~\cite{mcmc1,mcmc2} and a custom theory
for the derivation of the background. The numerical solutions of the field
equations (\ref{qq.19}) - (\ref{qq.21}) are derived with the Runge-Kutta
method. 

The analysis of the MCMC chains was performed with the GetDist
library\footnote{https://getdist.readthedocs.io/}~\cite{getd}.

For the observational data, we consider the following datasets OHD\&BAO,
SNe~Ia\&OHD\&BAO, where for SNe~Ia we assume the PP or the U3 catalogs. We
calculate the parameters which~maximize the likelihood $\mathcal{L}_{\max
}=\exp\left(  -\frac{1}{2}\chi_{\min}^{2}\right)  $, that is,%
\begin{equation}
\mathcal{L}_{\max}=\exp\left(  -\frac{1}{2}\chi_{OHD}^{2}\right)  \exp\left(
-\frac{1}{2}\chi_{BAO}^{2}\right)  ,
\end{equation}
or%
\begin{equation}
\mathcal{L}_{\max}=\exp\left(  -\frac{1}{2}\chi_{SNI_{a}}^{2}\right)
\exp\left(  -\frac{1}{2}\chi_{OHD}^{2}\right)  \exp\left(  -\frac{1}{2}%
\chi_{BAO}^{2}\right)  .
\end{equation}
We perform the constraints for the Quintom-Chiral model, for the $\Lambda$CDM
and the $w_0w_a$CDM model, which is used as a reference point. In the general case,
the free parameters for the Chiral-Quintom model form a vector space of
dimension seven, they are, $\left\{  H_{0},\Omega_{m0},r_{drag},x_{0},\xi
_{0},\lambda,\kappa\right\}  _{\text{CQ}}$, for the $\Lambda$CDM they form a
vector space of dimension three, that is, $\left\{  H_{0},\Omega_{m0}%
,r_{drag}\right\}  _{\Lambda\text{CDM}}$ and for the $w_0w_a$CDM model the parametric
space has dimension five, that is, $\left\{  H_{0},\Omega_{m0},r_{drag}%
,w_{0},w_{a}\right\}  _{\Lambda\text{CDM}}$. Due to the symmetries of the
dynamical system for the Chiral-Quintom\ model, we consider $\xi_{0}>0$, and
without loss of generality we assume $\lambda<0$, as long as $x,$and $\kappa$
can take positive and negative values. 

The priors considered in this study for the COBAYA interface are presented
in Table \ref{prior}.%

\begin{table}[tbp]
    \centering
    \caption{Priors adopted for the free parameters in the MCMC analysis for the Chiral-Quintom and $\Lambda$CDM models.}
    \label{prior}
    \setlength{\tabcolsep}{10pt}
    \renewcommand{\arraystretch}{1.20}
    \begin{tabular}{lccc}
        \toprule
        \textbf{Parameter} & \textbf{Chiral-Quintom} & \textbf{$\Lambda$CDM} & \textbf{$w_0w_a$CDM}  \\
        \midrule
        $\mathbf{H}_{0}$ & $[60,\,80]$ & $[60,\,80]$ & $[60,\,80]$ \\ 
        $\mathbf{\Omega}_{m0}$ & $[0.1,\,0.5]$ & $[0.1,\,0.5]$ & $[0.1,\,0.5]$ \\ 
        $\mathbf{r}_{\mathrm{drag}}$ & $[120,\,170]$ & $[120,\,170]$ & $[120,\,170]$ \\ 
        $\mathbf{x}_{0}$ (equiv.\ $\mathbf{w}_{0}$) & $[-0.9,\,0.9]$& --- & $[-1.1,\,-0.6]$ \\
        $\mathbf{\xi}_{0}$ (equiv.\ $\mathbf{w}_{a}$) & $[0.9]$ & --- & $[-1.1,\,0.5]$ \\ 
        $\mathbf{\lambda}$ & $[-10,\,0]$ & --- & --- \\ 
        $\mathbf{\kappa}$ & $[-150,\,0]$ & --- & ---\\ 
        \textbf{Constraint} & Eq.\ (\ref{qq.25}) & --- & ---\\
        \bottomrule
    \end{tabular}
\end{table}

The Akaike Information Criterion (AIC) \cite{AIC} is applied to compare the
statistical significance of the two models. Specifically, from the
$\mathcal{L}_{\max}$, for each dataset we calculate the $AIC$ parameter as
follows
\begin{equation}
AIC\simeq-2\ln\mathcal{L}_{\max}+2\mathcal{N},
\end{equation}
where $\mathcal{N}$ is the dimension of the parametric space.

AIC defines the statistical criterion to determine the preference over different models. More precisely, 

\begin{itemize}
    \item[-] if $\left\vert \Delta AIC\right\vert
=\left\vert AIC_{QC}-AIC_{\Lambda}\right\vert$. Specifically, for $\left\vert
\Delta AIC\right\vert <2$, the two models are statistically equivalent,
\item[-] if  $\left\vert \Delta AIC\right\vert <6$, there is weak evidence in favor of the
model with smaller $AIC$,
\item[-] if $\left\vert \Delta AIC\right\vert >6$ the model under exam appears disfavored significantly.  
\end{itemize}

\section{Discussion on cosmological Constraints}

We continue with the presentation of the observational constraints. To do so, after performing the analysis for the Chiral-Quintom model with generic parameters
$\kappa$ and $\lambda$, we can see in Table \ref{best1}, the best-fit 
parameters for the Chiral-Quintom model and for other two paradigms, namely the standard $\Lambda$CDM and $w_0w_a$CDM parametrization. 

Moreover, in Table \ref{chi1}, our statistical comparisons are thus performed among the three scenarios.

In particular, we can summarize below our main findings. 

\begin{figure}[h!]
\centering\includegraphics[width=0.5\textwidth]{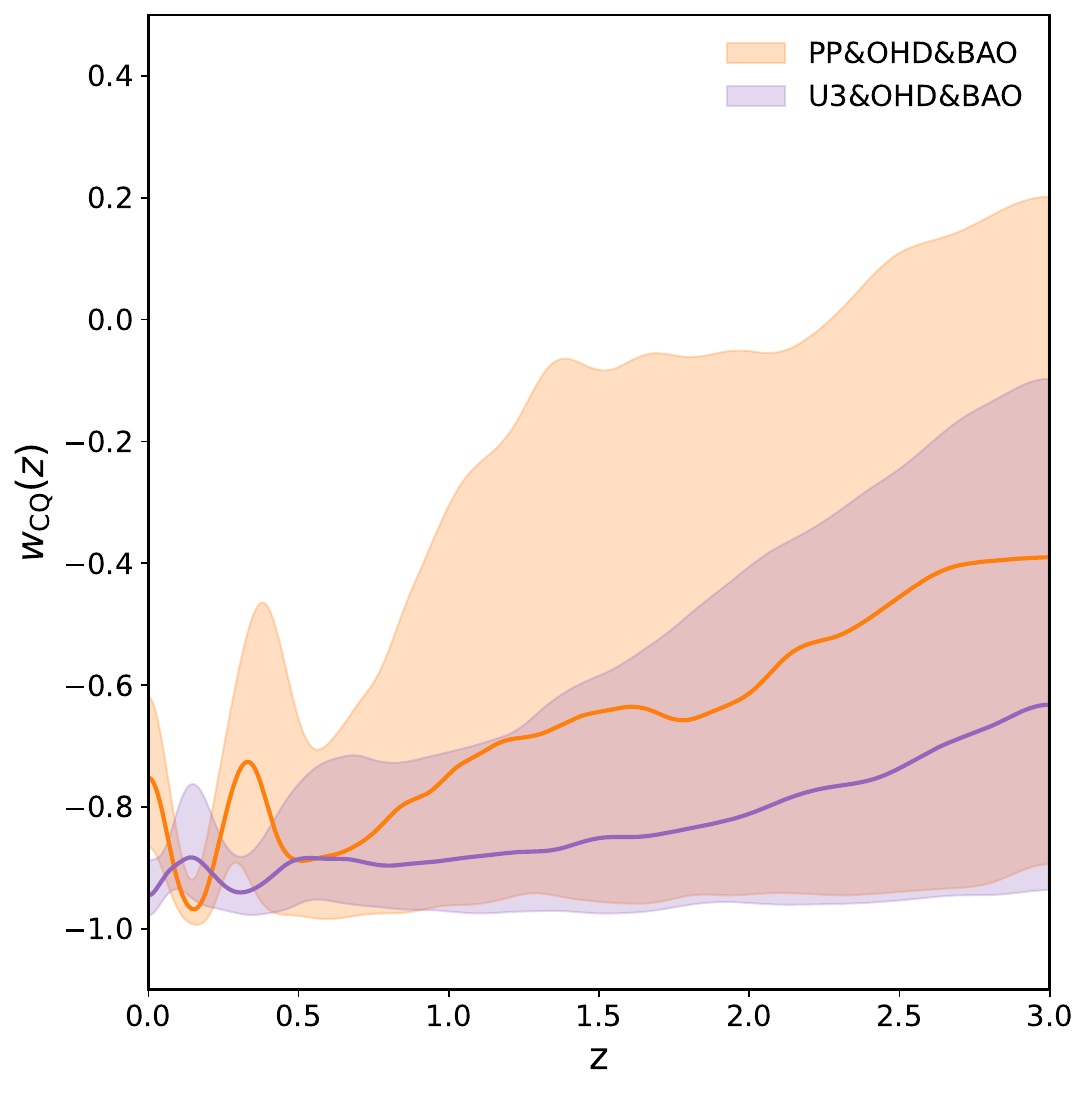}\caption{Evolution of
the dark energy equation of state parameter $w_{CQ}\left(  z\right)  $ for the
posterior variables of Table \ref{best1} within $1\sigma$. Solid lines are for
the mean values. We observe that the present acceleration phase is described
by the asymptotic solution where the two scalar fields contribute to the
cosmic fluid. }%
\label{fig2}%
\end{figure}

\begin{itemize}
    \item[-] {\bf PP\&OHD\&BAO}. The introduction of the SNe~Ia data from the PP catalog into the previous
dataset reveals that the model that fits the data with the maximum likelihood
value is the Chiral-Quintom model. As before, the comparison of the statistical
parameters with $\Lambda$CDM as the baseline model gives
$\Delta\chi_{\min}^{2}=-10.99$ and $\Delta AIC^{CQ}=-2.99$. Consequently,
according to Akaike’s scale, we conclude that the data provide weak evidence
in favor of the Chiral-Quintom model. Furthermore, the model fits the data
better than the CPL parametrization, with $\Delta\chi_{\min}^{2}=-7.6$, while the AIC gives
$\Delta AIC^{CQ}=-3.6$, meaning that Akaike’s scale again indicates weak
evidence in favor of the Chiral-Quintom model.

    \item[-] {\bf U3\&OHD\&BAO}. Finally, for the U3 catalog, we find that the Chiral-Quintom model is the one
that provides the maximum likelihood value. Moreover, when using $\Lambda$CDM
as the baseline model, the differences in the statistical parameters are
$\Delta\chi_{\min}^{2}=-12.49$ and $\Delta AIC^{CQ}=-4.49$. From this, we
conclude that according to the AIC the data provide strong evidence in favor
of the Chiral-Quintom model. Additionally, the model fits the data better than
the $w_0w_a$CDM model, with $\Delta\chi_{\min}^{2}=-4.51$. Nevertheless, due to the different
degrees of freedom, the data do not show a clear preference between the two
models. In the Appendix, specifically in Fig.~\ref{fig1}, we present the
$1\sigma$ and $2\sigma$ confidence regions for the free parameters of the
Chiral-Quintom model, whereas the dark energy equation of state
$w_{CQ}(z)$ as a function of redshift is shown in Fig.~\ref{fig2}, adopting
the mean values reported in Table \ref{best1}.
\end{itemize}

\begin{table}[tbp]
\centering
\caption{Numerical outcomes inferred from our model in comparison with the standard cosmological background and the $w_0w_a$CDM parametrization.}
\label{best1}
\begin{tabular}{lccccccc}
\toprule
 & $H_0$ 
 & $\Omega_{m0}$ 
 & $r_{\mathrm{drag}}$ 
 & $x_0\,(w_0)$
 & $\xi_0\,(w_a)$
 & $\lambda$
 & $\kappa$ \\
\midrule
\multicolumn{8}{c}{\textbf{PP \& OHD \& BAO}} \\
\midrule
\textbf{Chiral-Quintom} 
 & $67.6_{-1.7}^{+1.7}$ 
 & $0.238_{-0.036}^{+0.043}$ 
 & $147.7_{-3.5}^{+3.5}$ 
 & $-0.01_{-0.17}^{+0.17}$ 
 & $0.27_{-0.10}^{+0.14}$ 
 & $-3.79_{-0.89}^{+0.20}$ 
 & $-99_{-44}^{+22}$ \\
$\Lambda$\textbf{CDM}
 & $68.5_{-1.6}^{+1.6}$ 
 & $0.311_{-0.012}^{+0.012}$ 
 & $147.0_{-3.4}^{+3.4}$ 
 & -- & -- & -- & -- \\
\textbf{$w_0w_a$CDM}
 & $68.0_{-1.7}^{+1.7}$
 & $0.308_{-0.015}^{+0.022}$
 & $147.1_{-3.6}^{+3.2}$
 & $-0.893_{-0.059}^{+0.059}$
 & $-0.33_{-0.52}^{+0.43}$
 & --
 & -- \\
\midrule
\multicolumn{8}{c}{\textbf{U3 \& OHD \& BAO}} \\
\midrule
\textbf{Chiral-Quintom}
 & $67.1_{-1.7}^{+1.7}$
 & $0.279_{-0.023}^{+0.031}$
 & $147.5_{-3.4}^{+3.4}$
 & $-0.01_{-0.012}^{+0.064}$
 & $0.16_{-0.12}^{+0.064}$
 & $< -1.73$
 & $< -70.9$ \\
$\Lambda$\textbf{CDM}
 & $68.5_{-1.7}^{+1.7}$
 & $0.311_{-0.013}^{+0.011}$
 & $147.0_{-3.4}^{+3.4}$
 & -- & -- & -- & -- \\
\textbf{$w_0w_a$CDM}
 & $66.9_{-1.7}^{+1.7}$
 & $0.319_{-0.014}^{+0.021}$
 & $147.1_{-3.5}^{+3.1}$
 & $-0.790_{-0.070}^{+0.089}$ 
 & $< -0.472$
 & --
 & -- \\
\bottomrule
\end{tabular}
\end{table}

\begin{table}[tbp]
    \centering
    \small
    \caption{Statistical comparison of the Chiral-Quintom model with respect to the $\Lambda$CDM and $w_0w_a$CDM scenarios for different data combinations.}
    \label{chi1}
    \setlength{\tabcolsep}{5pt}
    \renewcommand{\arraystretch}{1.2}
    \begin{tabular}{llccc}
        \toprule
        Data set 
        & Reference model 
        & $\Delta\chi^{2}_{\min}$ 
        & $\Delta\mathrm{AIC}$ 
        & Akaike's scale \\
        \midrule
        \textbf{PP \& OHD \& BAO} 
        & $\Lambda$CDM 
        & $-10.99$ 
        & $-2.99$ 
        & Weak evidence in favor of $CQ$ \\
        \textbf{PP \& OHD \& BAO} 
        & $w_0w_a$CDM 
        & $-7.60$  
        & $-3.60$ 
        & Weak evidence in favor of $CQ$ \\
        \textbf{U3 \& OHD \& BAO} 
        & $\Lambda$CDM 
        & $-12.49$ 
        & $-4.49$ 
        & Strong evidence in favor of $CQ$ \\
        \textbf{U3 \& OHD \& BAO} 
        & $w_0w_a$CDM 
        & $-4.51$  
        & $-0.51$ 
        & Inconclusive \\
        \bottomrule
    \end{tabular}
\end{table}

\section{Final outlooks}\label{sec4}

In this work, we investigated the \textit{Chiral--Quintom} multi-scalar field model as a viable framework aiming to unify inflation and dark energy, and accordingly we propose it as a consistent candidate to explain late-time cosmic acceleration.

Specifically, we considered a simplified version of the two-field Lagrangian introduced in Refs.~\cite{sa1,anuni}, which naturally leads to a multi-scalar field sector with an explicitly nonminimal coupling structure. In particular, we assumed that the kinetic term of the second scalar field is nonminimally coupled to the first one through a mixing function controlled by the constant parameter $\kappa$. This choice introduces an additional degree of freedom that mediates the interaction between the two fields.

Focusing on the late-time behavior of the universe, we adopted an exponential potential to govern the scalar field dynamics. This potential can be interpreted as the asymptotic limit of a more general class of potentials and has been widely employed in the literature, see e.g.~\cite{Copeland:1997et,Ferreira:1997hj,Barreiro:1999zs}. Moreover, for small-field expansions, it reduces to a quadratic potential with higher-order corrections, as discussed for example in~\cite{delaMacorra:1999ff,Carloni:2023egi,Carloni:2025kev}.

Within this setup, the model predicts the emergence of \emph{two distinct} accelerated phases of expansion. The first corresponds to a standard slow-roll regime, analogous to a single field quintessence, where the dynamics is effectively driven by a unique scalar field. The second regime arises asymptotically, when both scalar fields become dynamically relevant, giving rise to a time-varying dark energy component characterized by an oscillatory behavior, in fulfillment with previous findings in the literature~\cite{sup0,sup1,sup2,sup3}.

Compared to the standard $\Lambda$CDM model, the \textit{Chiral--Quintom} framework introduces four additional free parameters: the two exponential indices $\lambda$ and $\kappa$, and two initial conditions associated with the evolution of the scalar fields. Accordingly, we remarked that this condition makes the model naturally comparable to the $w_0 w_a$CDM parametrization, i.e., the phenomenological scheme used by the DESI collaboration to emphasize the discrepancies w.r.t. the $\Lambda$CDM case.

To constrain the aforementioned parameters, we employed a comprehensive set of late-time cosmological observations, including OHD, SNe~Ia, and BAO measurements from the DESI DR2 release. Our analysis shows that the hierarchy $\kappa / \lambda \gg \lambda$ is statistically favored, ensuring that the present value of the dark energy equation of state parameter remains close to $w_{\mathrm{DE}} \approx -1$. Furthermore, we find that $\lambda$ is constrained to $\lambda < -\sqrt{2}$, a result that plays an important  role in shaping the oscillatory nature of the effective dark energy fluid. Remarkably, this condition also suggests that the model can support an early inflationary phase, dominated by a slow-roll behavior where only the quintessence field contributes to the total energy budget, as a consequence of the fact that, at large redshifts $z$, one finds $\xi(z)\rightarrow 0$.

Future works will shed light on alternative multifield approaches capable of unifying inflation and dark energy. This class of models may provide a key to clarify why the universe seems to exhibit two distinct accelerated phases, each characterized by the dominance of a negative equation of state, albeit with quite different energy scales. Further, we will investigate more deeply the impact of DESI data on multifield physics, also in view of additional forthcoming data sets. In this respect, the characterization of our model and/or its extensions will be compared with cosmic microwave background observations to assess both early-time stability and possible observational signatures.

\begin{acknowledgments}
AP acknowledges hospitality hosted by the University of Camerino while
this work has been carried out. In particular, AP is also thankful to OL for the invitation to join his research group during his stay at the University of Camerino. AP \& GL thank the support of
Vicerrector\'{\i}a de Investigaci\'{o}n y Desarrollo Tecnol\'{o}gico (Vridt)
at Universidad Cat\'{o}lica del Norte through N\'{u}cleo de Investigaci\'{o}n
Geometr\'{\i}a Diferencial y Aplicaciones, Resoluci\'{o}n Vridt No - 096/2022.
This work was partially supported by Proyecto Fondecyt Regular Folio 1240514,
Etapa 2025. TM is grateful to Paulo M. S\'a for the productive period spent at his institute, during which the basic ideas behind this work were partially proposed. OL acknowledges financial support from the Fondazione  ICSC, Spoke 3 Astrophysics and Cosmos Observations. National Recovery and Resilience Plan (Piano Nazionale di Ripresa e Resilienza, PNRR) Project ID CN$\_$00000013 "Italian Research Center on  High-Performance Computing, Big Data and Quantum Computing"  funded by MUR Missione 4 Componente 2 Investimento 1.4: Potenziamento strutture di ricerca e creazione di "campioni nazionali di R$\&$S (M4C2-19 )" - Next Generation EU (NGEU)
GRAB-IT Project, PNRR Cascade Funding
Call, Spoke 3, INAF Italian National Institute for Astrophysics, Project code CN00000013, Project Code (CUP): C53C22000350006, cost center STI442016.   
\end{acknowledgments}

\newpage
\begin{appendix}

\section*{Contour plots and numerical results}

In this appendix, we report our contours, computed up to the $2\sigma$ confidence level. The likelihoods are reported on the right, denoting mild convergence. The need of additional data points will be faced in future analyses, adding more data points, as stated in the perspectives.

   \begin{figure}[h!]
\centering\includegraphics[width=0.8\textwidth]{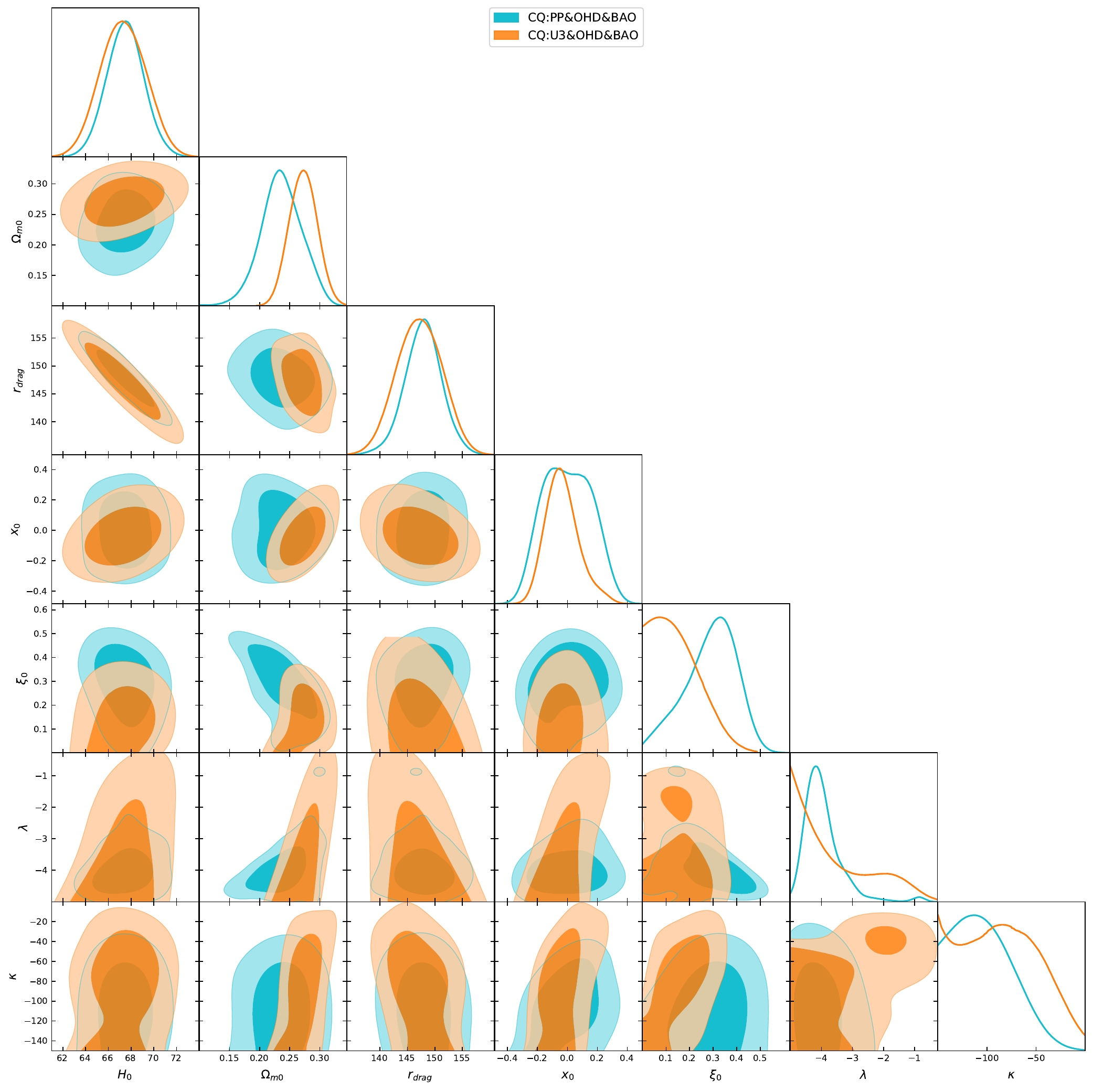}\caption{Contour plots based on $1\sigma$ and $2\sigma$ confidence levels for the Chiral-Quintom model.}%
\label{fig1}%
\end{figure} 
\end{appendix}

\end{document}